\documentclass[1p]{elsarticle}
\usepackage{graphicx}
\usepackage{amsmath}
\usepackage{amssymb}
\usepackage{bm}
\usepackage{epstopdf}
\begin{document}

\begin{frontmatter}

\title{ Coulomb impurity problem of graphene in magnetic fields}
\author{  S. C. Kim and S. -R. Eric Yang\footnote{Corresponding author. Tel.:+82 2 3290 3100.\\ E-mail address: eyang812@gmail.com (S. -R. Eric Yang).}}
\address{Physics Department, Korea University, Seoul, 136-713, Korea}

\begin{abstract}
Analytical solutions of the Coulomb impurity problem of graphene in
the absence of a magnetic field  show that when the dimensionless
strength of the Coulomb potential $g$ reaches a critical value  the
solutions become supercritical with imaginary eigenenergies.
Application of a magnetic field is a singular perturbation, and  no
analytical solutions are known except at a denumerably infinite set
of magnetic fields. We find solutions of this problem by numerical
diagonalization of large Hamiltonian matrices. Solutions are
qualitatively different from those of zero magnetic field.  All
energies are discrete and no complex energies allowed.   We have
computed the finite-size scaling function of the probability density
containing  s-wave component of Dirac wavefunctions. This function
depends on the coupling constant, regularization parameter, and the
gap.  In the limit of vanishing regularization parameter our
findings are  consistent with the expected values exponent $\nu$
which determines of the asymptotic behavior of the wavefunction near
$r=0$.
\end{abstract}

%\begin{keyword}
%\PACS
%\end{keyword}

\end{frontmatter}

\section {Introduction}

States of relativistic electrons in the three dimensional Coulomb
impurity problem can become supercritical when the charge of the
nucleus becomes sufficiently large\cite{Rei}.  Recently similar
problem has attracted a lot of attention in two-dimensional
graphene.   The Hamiltonian\cite{Mac0,Neto} is
\begin{eqnarray}
 H=
v_F \vec{\sigma}\cdot({\vec p}+\frac{e}{c}{\vec A}
)-\frac{Ze^2}{\epsilon r}+\Delta \sigma_z,
\end{eqnarray}
where  $\vec{\sigma}=(\sigma_x,\sigma_y)$ and $\sigma_{z}$ are Pauli
spin matrices ($\vec{p}$ is two-dimensional momentum and $\epsilon$
is the dielectric constant). A magnetic field $\vec{B}$ is applied
perpendicular to the two-dimensional plane and the vector potential $\vec A$ is
given in a symmetric gauge. In the presence of a finite mass gap
$\Delta$ a new term $\Delta \sigma_z$ is added to the Hamiltonian.
 Angular momentum $J$ is a
good quantum number and wavefunctions of eigenstates have the form
\begin{equation}
\Psi^{J}(r,\theta)=e^{i(J-1/2)\theta}\left(
\begin{array}{c}
\chi_{A}(r)\\
\chi_{B}(r)e^{i\theta}
\end{array}
\right).
\end{equation}
It consists of  A and B radial wavefunctions $\chi_A(r)$ and
$\chi_B(r)$ with channel angular momenta $J-1/2$  and $J+1/2$,
respectively. The half-integer angular momentum quantum numbers have
values $J=\pm1/2,\pm3/2,\cdots$. In this  paper we will consider
only states that have a s-wave component, namely states with $J=\pm
1/2$.

The dimensionless coupling constant of the Coulomb potential is
\begin{eqnarray}
g=\frac{Ze^2}{\epsilon\hbar v_F}.
\label{coupling}
\end{eqnarray}
In the absence of a magnetic field and zero mass gap subcritical and
supercritical regimes separate at the critical coupling constant
$g_c=1/2$\cite{Per,Gam0}. In subcritical regime $g<1/2$ no natural
length scale exists   since the Bohr radius is undefined when
$\Delta=0$, and no boundstates exist and only scattering states
exist (when $\Delta\neq0 $ the effective Bohr radius is given by
$\lambda=\frac{1}{g}\frac{\hbar v_F}{\Delta}$). This is in a quite
contrast to the Coulomb impurity problem of an ordinary
two-dimensional electron in magnetic fields with the Bohr radius
$\frac{\epsilon\hbar^2}{m e^2}$\cite{Mac} ($m$ is the electron
mass). In the supercritical regime $g>1/2$ a spurious effect of the
fall into the center of potential appears\cite{Landau, Rei}: the
solution diverges in the limit $r\rightarrow 0$ and  exhibits pathological
oscillations near $r=0$.

This spurious effect can be circumvented by regularizing the Coulomb
potential with  a length scale $R$\cite{com}, and physically
acceptable complex energy states (quasi-stationary levels)
appear\cite{Rei}.  A resonant state with angular momentum $J=1/2$
has a complex energy $E$ that depends on $g$\cite{Gam0}
\begin{eqnarray}
\frac{E}{E_R}=-(1.18+0.17i)e^{\frac{-n\pi }{\sqrt{g^2-g_c^2}}}
\label{Bequal_zero}
\end{eqnarray}
for $g-g_c\ll 1$ and $\Delta=0$, where  the characteristic energy
scale associated with the length scale $R$ is
\begin{eqnarray}
E_R=\hbar v_F/R.
\end{eqnarray}
In the limit $R\rightarrow 0$ the size of the wavefunction goes to
zero and the real part of the energy {\it diverges} toward
$-\infty$, see Eq.(\ref{Bequal_zero}). These results indicate that
the electron falls to the center of potential.  In the presence of a
gap Gamayun et al.\cite{Gam0} find that the critical coupling constant for the
angular momentum $J=1/2$ is
\begin{eqnarray}
g_c(\Delta,E_R)=\frac{1}{2}+\frac{\pi^2}{\log^2(c\frac{\Delta
}{E_R})}, \label{critical}
\end{eqnarray}
where $c\approx 0.21$.  Complex energies appear for $g>g_c$.
According to this result the presence of a mass gap does not change
the critical value $g_c=0.5$ in the limit $R\rightarrow 0$.

The aim of the present paper is to  investigate the Coulomb impurity
problem of massless Dirac electrons in the {\it presence} of a
magnetic field. In this problem the meaning of subcritical and
supercritical states is  ill-defined. This is because  {\it no
complex energy solutions} (resonances) are possible in the Coulomb
impurity problem in magnetic fields: the effective potential does
not allow resonant states since the vector potential diverges while
the Coulomb potential goes to zero in the limit $r\rightarrow
\infty$\cite{gia,Rec}.   The $B\rightarrow 0$ limit is
singular\cite{Berry,Ben} since real energies of $B\neq 0$ change
into complex energies for $g>1/2$ and $\Delta=0$. It is unclear how
the wavefunctions of subcritical and supercritical regions of $B=0$
change when $B\neq 0$.    Ho and Khalilov\cite{Ho} have provided
exact solutions below $g<0.5$ at a denumerably infinite set of
magnetic fields when $\Delta\neq 0$ and $R=0$. However, as far as we
know, no  analytical solutions are known for general values of $g$,
$R$, $\Delta$, and $B$, especially for $g>1/2$.   Zhang et
al.\cite{Zh} have investigated this problem using the WKB method for
$\Delta=0$ and $g<1/2$.  They also argue that $\nu=\sqrt{\frac{1}{4}-g^2}$ for
$g<1/2$.

Here we find solutions by diagonalizing numerically large
Hamiltonian matrices using the graphene Landau level (LL) states as the basis states for
various values of $\Delta$, $g$, and $R$. The dimension of
Hamiltonian matrix $N_c$ acts as a cutoff parameter, and is related
to the regularization parameter of the Coulomb potential
$R$\cite{com01}
\begin{eqnarray}
R\sim \ell\pi\sqrt{\frac{2}{N_c}}, \label{reg}
\end{eqnarray}
where $\ell$ is the magnetic length.  We find that all the states
are discrete and no complex energies allowed.  The obtained
eigenstates $|\Psi^J\rangle$ with angular momentum $J$ can be
labeled additionally  by the LL index $N$: $|\Psi_N^J\rangle$. The
corresponding eigenvalues are denoted by $E_N^J$.   The computed
energy spectrum is consistent with available analytical results. Our
{\it finite-size scaling} analysis\cite{Golden} shows that the value
of the probability density value of the state
$|\Psi_0^{-1/2}\rangle$ at $r=R$ is described by the following
scaling function
\begin{eqnarray}
\frac{1}{\ell^2|\Psi_{0}^{-1/2}(R)|^2}=f\Big(g,\frac{1}{N_c},\frac{\Delta}{E_M}\Big),
\label{inv-den}
\end{eqnarray}
where
\begin{eqnarray}
E_M=\frac{\hbar v_F}{\ell}
\end{eqnarray}
is the energy scale associated with graphene LLs. We have also
computed electronic wavefunctions as a function of $r$ for various
values of $g$. These are main results of our work. The wavefunction
of s-wave component behaves as $\frac{1}{r^{\nu}}$ near $r=0$ in the
limit $R\rightarrow 0$. The exponent $\nu$ is determined through
data collapse of numerical results. When $g<1/2$  we find that the
exponent is $\nu<1/2$. For $g>1/2$ we find $\nu=1/2$, independent of
$g$ and $\Delta$.  In the limit $R\rightarrow 0$ our scaling results
are thus consistent with the known results\cite{Gam0,Zh}.

This paper is organized as follows.  In Sec.2 a Hamiltonian matrix
method is described.  Scaling properties of wavefunctions are given
in Sec.3.   The obtained eigenvalues and eigenstates are given in
Secs.4 and 5. In the last Sec.6 we give conclusions and
discussions.

\section {Hamiltonian matrix}

We compute eigenstates and eigenvalues by solving the Hamiltonian
matrix.   The  Hamiltonian matrix elements are constructed  using
graphene Landau level states as the basis states. We divide the
Hilbert space into subspaces of angular momentum
$J=|n|-m-\frac{1}{2}$. In each Hilbert subspace the eigenstates can
be written as a linear combination
\begin{eqnarray}
\Psi_N^J(\vec{r})=\sum_n C_n \psi_{n,m}(\vec{r}), \label{eq:expan}
\end{eqnarray}
where the basis vectors $\psi_{n,m}(r)$ are  the LL states of
graphene with angular momentum $J$ (This linear combinations is
expected to be accurate for $r>R$ because of the cutoff $N_c$). Note
that this method is valid only when $B\neq0$. The basis states are
given by
\newcommand{\ud}{\mathrm{d}}
\begin{eqnarray}
\psi_{n,m}(\vec{r})=c_{n}\left(\begin{array}{c}-\textrm{sgn}(n)i\phi_{|n|-1,m}(\vec{r})\\
\phi_{|n|,m}(\vec{r})\end{array}\right),\label{twocomp}\label{eq:spinor}
\end{eqnarray}
where $c_n=1$ for $n=0$ and $1/\sqrt{2}$ otherwise.  Their energies
are the LL energies $E_n=\textrm{sgn}(n)E_M\sqrt{2|n|}$ with the
wavefunctions\cite{Yosi}
\begin{eqnarray}
\phi_{n,m}(\vec{r})&=&A_{n,m}\exp\left(i(n-m)\theta-\frac{r^2}{4\ell^2}\right)\left(\frac{r}{\ell}\right)^{|m-n|}\nonumber\\
&\times&L_{(n+m-|m-n|)/2}^{|m-n|}\left(\frac{r^2}{2\ell^2}\right),\label{basis}
\end{eqnarray}
where $L_p^{\alpha}(z)$ is the Laguerre polynomial.  Here the
normalization factor is
\begin{eqnarray}\label{Anm}
A_{n,m}=\frac{1}{\ell}\left(2\pi\,2^{\alpha}\frac{\Gamma\big[{\beta}
+{\alpha}+1\big]}{\beta!}\right)^{-1/2},
\end{eqnarray}
where $\alpha=|m-n|$ and $\beta=(n+m-\alpha)/2$.

\subsection{Without mass term}

To perform extensive computation efficiently it is important to
evaluate the matrix elements analytically.  In units of the energy
scale of LLs $E_M$ the matrix elements of the kinetic operator are
diagonal with respect to the basis states and are
\begin{eqnarray}\label{Anm}
H_{n,n}=\textrm{sgn}(n)\sqrt{2|n|}.
\end{eqnarray}
The impurity potential conserves the angular momentum quantum number
$J$ and the impurity potential matrix elements in the Hilbert
subspace $J$ are
\begin{eqnarray}
&&V_{n,n'}=\langle\psi_{n,m}|\frac{Ze^2}{\epsilon rE_M}|\psi_{n',m'}\rangle=2\pi g c_nc_{n'}\times\nonumber\\
&&\bigg[\textrm{sgn}(nn')2^{\alpha_2-1/2}A_{\alpha_2,\beta_2}A_{\alpha_2,\beta_2'} I_{\beta_2,\beta_2'}(\alpha_2-1/2,\alpha_2,\alpha_2)\nonumber\\
&&+2^{\alpha_1-1/2}A_{\alpha_1,\beta_1}A_{\alpha_1,\beta_1'}
I_{\beta_1,\beta_1'}(\alpha_1-1/2,\alpha_1,\alpha_1)\bigg],\nonumber\\
\end{eqnarray}
where  $\alpha_1=|J-1/2|$, $\beta_1=\frac{2|n|-J-3/2}{2}$,
$\beta_1'=\frac{2|n'|-J-3/2}{2}$, $\alpha_2=|J+1/2|$,
$\beta_2=\frac{2|n|-J-1/2}{2}$, and
$\beta_2'=\frac{2|n'|-J-1/2}{2}$. Note that
\begin{eqnarray}\label{IntegralOfProductOfLaguerrePolynomials}
    &&I_{n,m}(\mu,\alpha,\beta)\nonumber\\
    &&
    =\Gamma(\mu+1)\frac{(\alpha+1)_m (\beta-\mu)_n}{n!
    m!}\nonumber\\
    &&\,\,\,\,\,\times\,_3F_2(-m,\mu+1,\mu-\beta+1;\alpha+1,\mu-\beta+1-n;1),\nonumber\\
\end{eqnarray}
where $_3F_2(a_{1},a_{2},a_{3};b_{1},b_{2};z)$ is the generalized
hypergeometric function and $(a)_n=\Gamma(a+n)/\Gamma(a)$. Note that
the dimensionless Hamiltonian matrix elements depend on the coupling
constant $g$, which is independent of $B$.

\subsection{With mass term}

In the presence of the  mass term\cite{Berry1} $J$ is still a good
quantum number. Using the orthogonality of Laguerre polynomials
\begin{eqnarray}
\int_{0}^{\infty}x^{\alpha}e^{-x}L_{m}^{\alpha}(x)L_{n}^{\alpha}(x)=\frac{\Gamma (n+\alpha+1)}{n!}\delta_{m,n},
\end{eqnarray}
we find that the matrix elements of the mass term can be written as
\begin{eqnarray}
\Delta_{n,n'}=\frac{\Delta
}{E_M}\langle\psi_{n,m}|\sigma_z|\psi_{n',m'}\rangle=-\frac{\Delta
}{E_M}\delta_{n,-n'}.
\end{eqnarray}
Here  the mass term is measured in units of $E_M$.  Note that
these dimensionless Hamiltonian matrix elements depend on $B$
through $\frac{\Delta}{E_M}$.

\section{Scaling properties }

Our numerical results for the state $|\Psi_0^{-1/2}\rangle$ can be
approximated by the following ansatz of the B-component of the
radial wavefunction for $r\gtrsim 0$:
\begin{eqnarray}\label{wavescal}
\chi_{B}(r)=\left\{\begin{array}{cc} A'r^{
-\frac{1}{2}+\sqrt{\frac{1}{4}-g^2}}(1+B' r+\cdots) &
 ,\ \ g<1/2
\\ C'r^{-1/2}(1+D' r+\cdots) & , \ \ g>1/2.
\end{array}\right.\nonumber\\
\end{eqnarray}
This wavefunction component is of s-wave. In the limit $r\rightarrow
0$ the B-component of the radial wavefunction behaves as
$\frac{1}{r^{\nu}}$, where $\nu=\frac{1}{2}-\sqrt{\frac{1}{4}-g^2}$
for $g<1/2$ and $\nu=1/2$ for $g>1/2$. The other wavefunction
component (A-component) goes to zero in the limit $r\rightarrow 0$
and can be ignored. The constants $A'$, $B'$, $C'$, and $D'$ depend
on the scaling variables. This wavefunction leads to the following
scaling ansatz for the inverse probability density
$\frac{1}{\ell^2|\Psi_{0}^{-1/2}(R)|^2}$ at $r=R$:
\begin{eqnarray}
&&h\Big(g,\frac{1}{N_c},\frac{\Delta}{E_M}\Big)\nonumber\\
&=&\left\{\begin{array}{cc} A\Big(\frac{1}{N_c}\Big)
^{\frac{1}{2}-\sqrt{\frac{1}{4}-g^2}}\Big(1-\frac{B}{N^{1/2}_c}+\cdots\Big)
& , \ \ g<1/2
\\ C \Big(\frac{1}{N_c}\Big) ^{1/2}(1-\frac{D}{N^{1/2}_c}+\cdots) &
, \ \ g>1/2.
\end{array}\right.\nonumber\\
\label{main}
\end{eqnarray}
The first terms are dominant and the second terms are corrections.
This scaling ansatz will be tested against the numerical scaling
results  $f\Big(g,\frac{1}{N_c},\frac{\Delta}{E_M}\Big)$ in Sec.4.

\section{Results of eigenstates }

We employ our matrix diagonalization method to investigate how the
wavefunctions change as the coupling constant changes. We employ
large Hamiltonian matrices of  dimension $N_c$. Since $N_c$ cannot
taken to be infinitely large we use a scaling analysis
to extract the relevant result for $N_c=\infty$ from the result of
finite-size matrices.

\subsection {$\Delta=0$}

\begin{figure}[!hbpt]
\begin{center}
\includegraphics[width=0.8\textwidth]{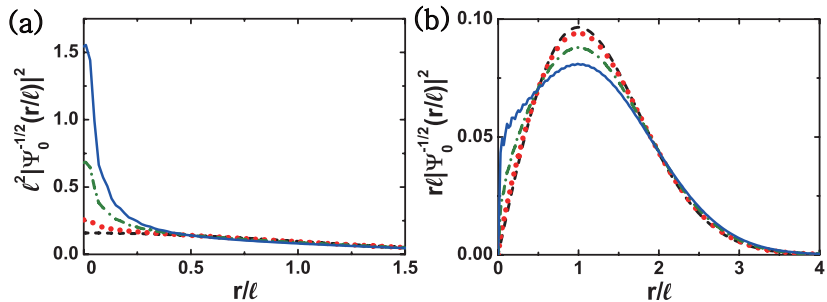}
\caption{(a) Computed values of  the probability density
$\ell^2|\Psi_0^{-1/2}(r/\ell)|^2$ in the absence of a gap for $N_c=8501$. The values of $g$ are  $0$ (dashed), $0.3$
(dotted), $0.5$ (dashed-dot), and $0.6$ (solid) ($\Delta=0$).
Corresponding energies are $0$, $-0.382$, $-0.663$, and $-0.835$ in
units of $E_M$.   In the absence of the Coulomb potential the
probability density (dashed line) at $r=0$ is $0.16$, which is much
smaller than the corresponding value in the presence of the Coulomb
potential. (b)  Plot of the radial probability density
$r\ell|\Psi_0^{-1/2}(r/\ell)|^2$. The values of $g$ are $0$ (dashed),
$0.3$ (dotted), $0.5$ (dashed-dot), and $0.6$ (solid) ($\Delta=0$).
The dimension of the Hamiltonian matrix is
$N_c=8501$.}\label{fig:jump}
\end{center}
\end{figure}

\begin{figure}[!hbpt]
\begin{center}
\includegraphics[width=0.8\textwidth]{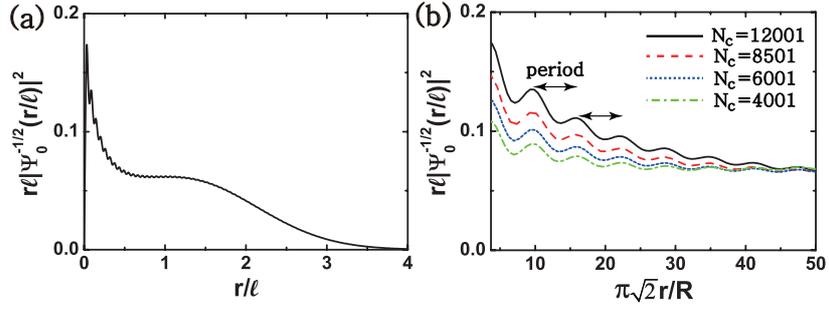}
\caption{ (a) Radial probability density $r\ell
|\Psi_0^{-1/2}(r/\ell)|^2$ for $N_c=12001$. Here $g=0.7$ and
$\Delta=0$. (b) Blow-up of the radial probability density $r\ell
|\Psi_0^{-1/2}(r/\ell)|^2$ near $r\gtrsim R$ is plotted as a
function of $\frac{r}{R}\pi\sqrt{2}$ for different values of
$N_c=12001, 8501, 6001, 4001$. }\label{fig:osc}
\end{center}
\end{figure}

Let us first calculate probability density
$\ell^2|\Psi_0^{-1/2}(r/\ell)|^2$ for $\Delta/E_M=0$. As the
coupling constant increases the probability density concentrates
near the center of the Coulomb potential, see Fig.\ref{fig:jump}(a).
It is more convenient to plot the radial probability density
 $(r/\ell)\ell^2 |\Psi_0^{-1/2}(r/\ell)|^2$ instead, see
Fig.\ref{fig:jump}(b) and Fig.\ref{fig:osc}(a):  we see that for
$g>1/2$  the value  of the radial probability density jumps nearly discontinuously near $r=R$.
This jump  becomes more sharper in the limit $R\rightarrow 0$
or $1/N_c \rightarrow 0$. This is consistent with the scaling
ansatz:  the radial wavefunction diverges as $\chi_B(r)\sim
r^{-1/2}$ near $r=0$, see Eq.(\ref{wavescal}). However, the
wavefunction is  {\it normalizable}. Fig.\ref{fig:osc}(b) replots
the radial probability density for
$g=0.7$ as a function of $\frac{r}{R}\pi\sqrt{2}$.   We see the
curves with different values of $N_c$ all have the same period,
approximately equal to $R$. The wavefunctions display stronger
oscillations with period $R$ in comparison to those of smaller
values of $g$.  These are Friedel-type oscillations originating from
an abrupt termination of the number of terms in the linear
combination of the eigenstates, see Eq.(\ref{eq:expan}).

\begin{figure}[!hbpt]
\begin{center}
\includegraphics[width=0.4\textwidth]{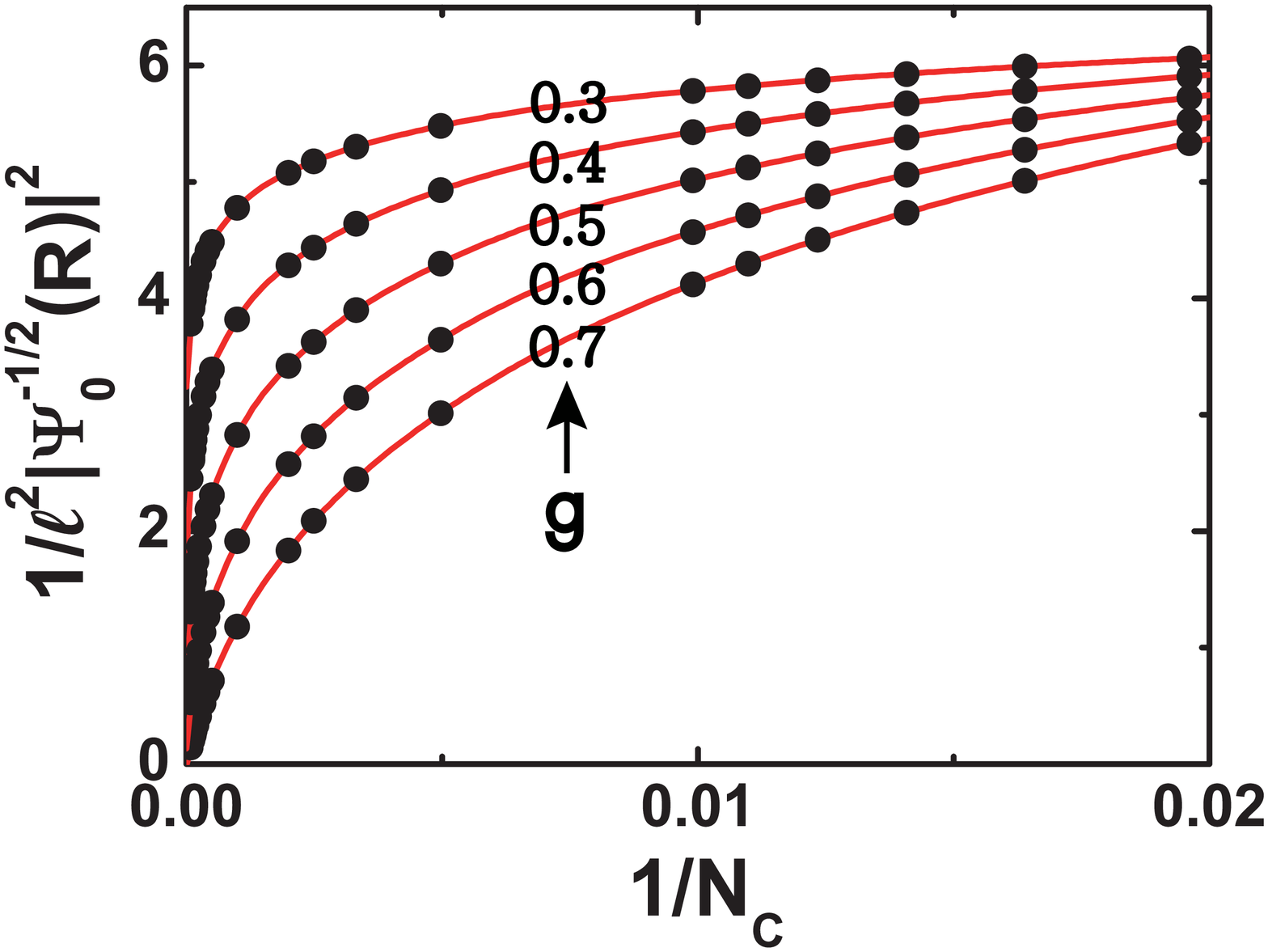}
\caption{Inverse probability density as a function of $1/N_c$
 for $g=0.3, 0.4, 0.5, 0.6, 0.7$.  The mass gap is $\Delta=0$.}\label{fig:invp}
\end{center}
\end{figure}

\begin{figure}[!hbpt]
\begin{center}
\includegraphics[width=0.4\textwidth]{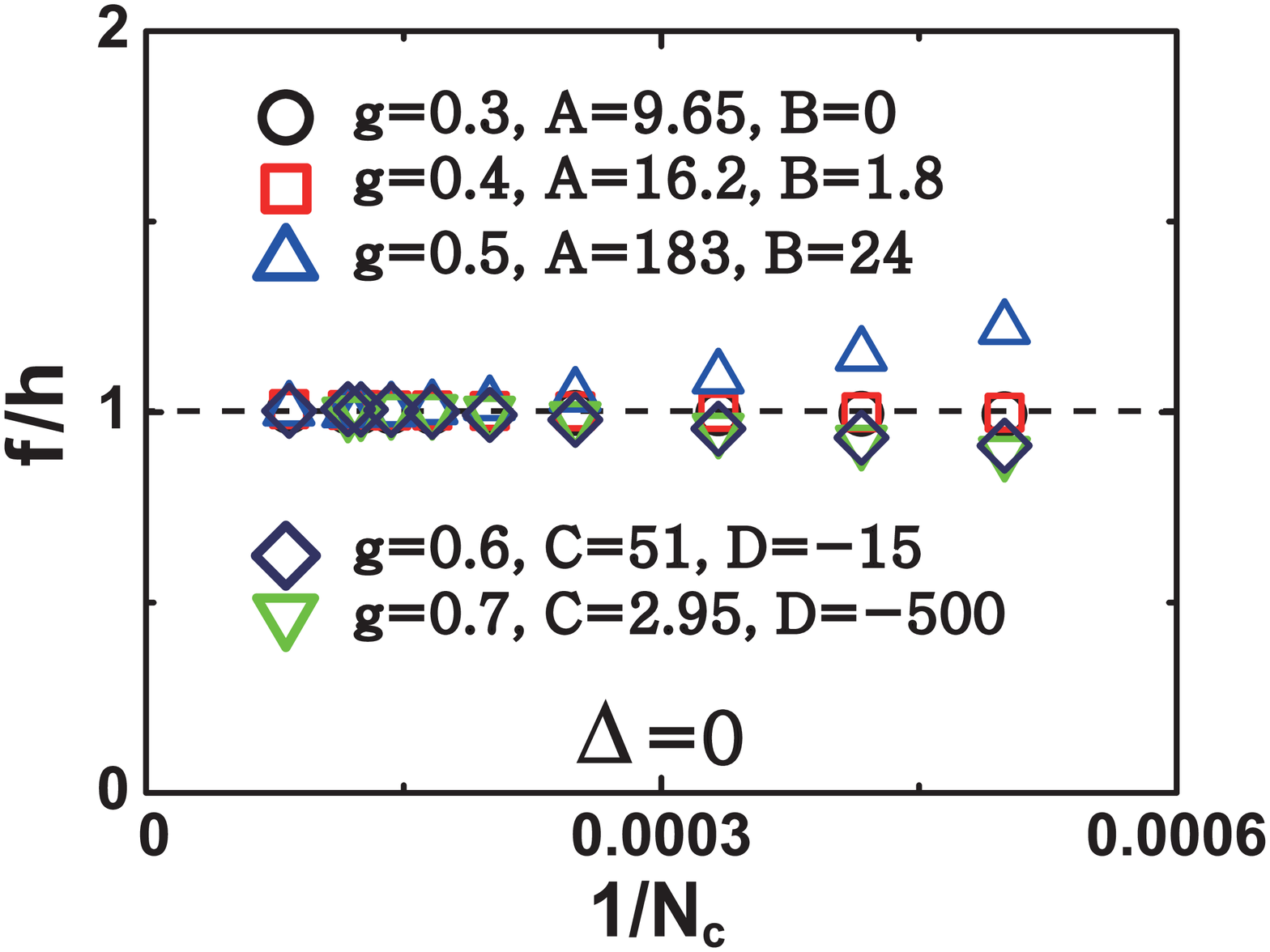}
\caption{Data collapse of the ratio $f/h$ in the limit
$1/N_c\rightarrow 0$ for  $\Delta=0$.
Here the dimension of the Hamiltonian matrix takes values $N_c=12001, 8501,
8001, 7001, 6001, 5001, 4001, 3001, 2401, 2001$.}\label{fig:delta0}
\end{center}
\end{figure}

Fig.\ref{fig:invp} displays the dimensionless inverse probability
density $f\Big(g,\frac{1}{N_c},0\Big)$ for various values of the
coupling constant.  Numerical result $f\Big(g,\frac{1}{N_c},0\Big)$
and the approximate scaling ansatz $h\Big(g,\frac{1}{N_c},0\Big)$ of
Eq.(\ref{main}) should agree when $\frac{1}{N_c}\ll 1$:
\begin{eqnarray}\label{Anm}
\lim_{\frac{1}{N_c}\rightarrow
0}\frac{f\Big(g,\frac{1}{N_c},0\Big)}{h\Big(g,\frac{1}{N_c},0\Big)}
= 1.
\end{eqnarray}
This is verified by the data collapse shown in Fig.\ref{fig:delta0}.
It confirms  our scaling ansatz given in Eq.(\ref{wavescal}).

\begin{figure}[!hbpt]
\begin{center}
\includegraphics[width=0.8\textwidth]{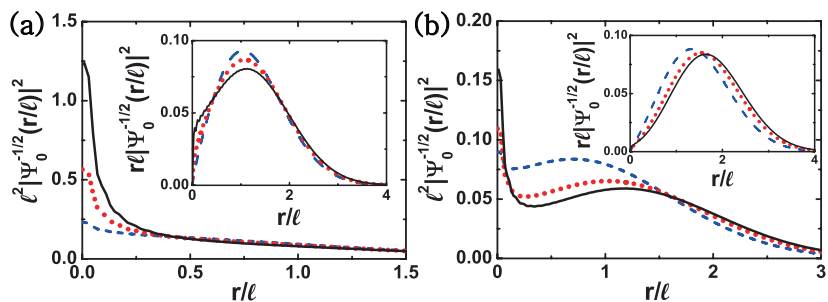}
\caption{(a) Probability density $\ell^{2}|\Psi_{0}^{-1/2}(r/\ell)|^2$ as a function
$r/\ell$. Values are  $g=0.3$ (dashed),   $0.5$ (dotted), and $0.6$
(solid). Corresponding energies are $-0.475$, $-0.741$, and $-0.897$
in units of $E_M$. Parameters are $J=-1/2$, $\Delta=0.1E_M$, and
$N_c=8501$. Inset: radial probability density
$r\ell|\Psi_{0}^{-1/2}(r/\ell)|^2$. (b) Probability density
$\ell^{2}|\Psi_{0}^{-1/2}(r/\ell)|^2$ as a function $r/\ell$. Values are $g=0.3$
(dashed),   $0.5$ (dotted), and  $0.6$ (solid). Corresponding
energies are $-1.327$, $-1.508$, and $-1.592$ in units of $E_M$.
Parameters are $J=-1/2$, $\Delta=E_M$,  and $N_c=8501$.   Inset:
radial probability density $r\ell|\Psi_{0}^{-1/2}(r/\ell)|^2$.
}\label{fig:prob2}
\end{center}
\end{figure}

\subsection{$\Delta\neq 0$}

\begin{figure}[!hbpt]
\begin{center}
\includegraphics[width=0.4\textwidth]{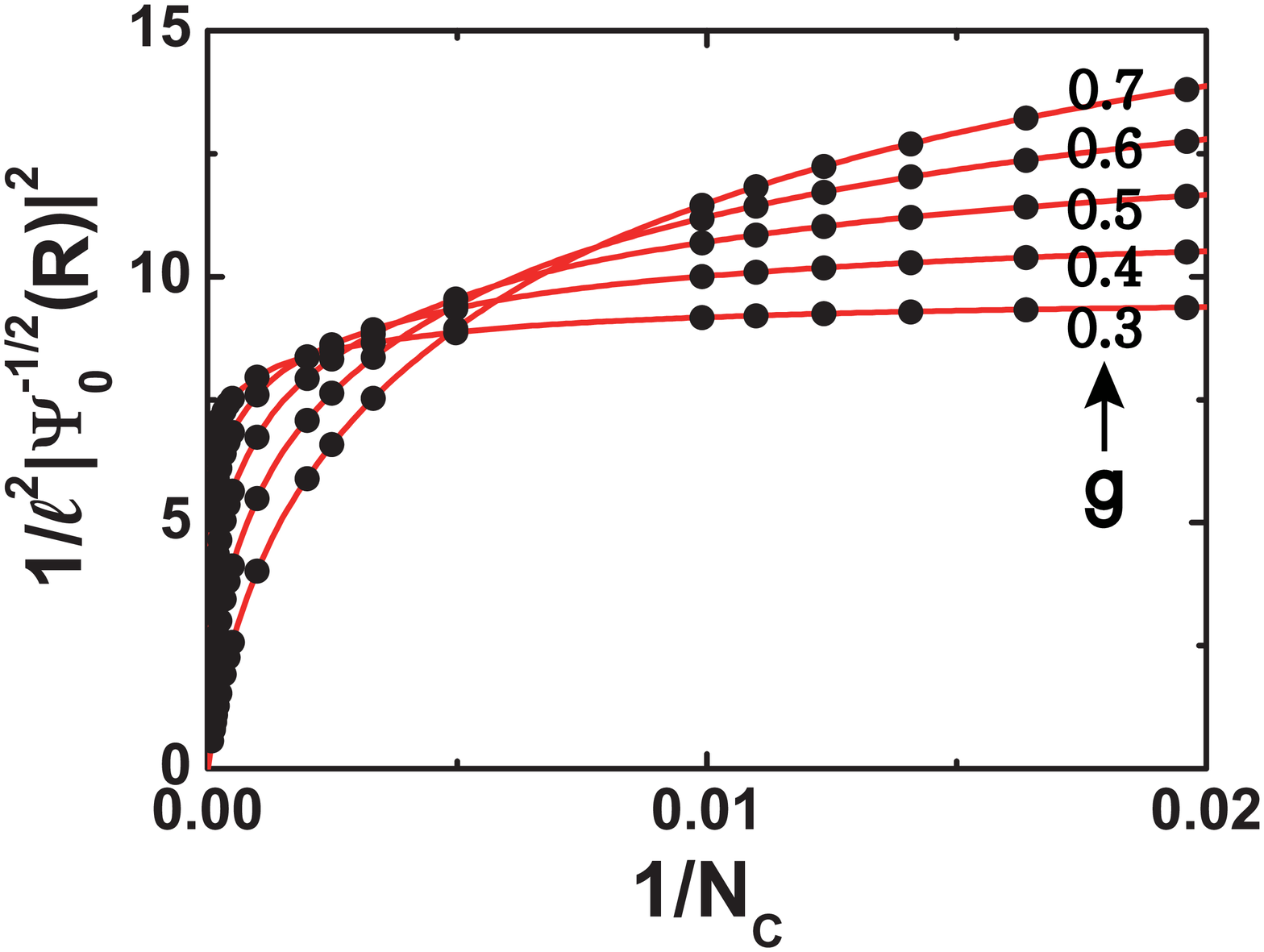}
\caption{Inverse probability density as a function of $1/N_c$
 for $g=0.3, 0.4, 0.5, 0.6, 0.7$.  The mass gap is $\Delta=0.5E_M$.}\label{fig:invp2}
\end{center}
\end{figure}

\begin{figure}[!hbpt]
\begin{center}
\includegraphics[width=0.4\textwidth]{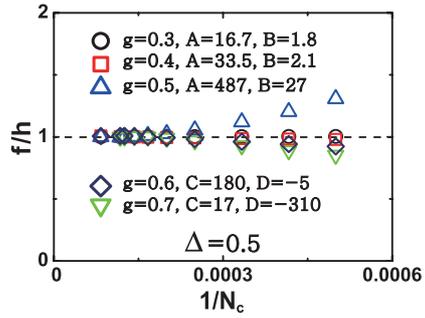}
\caption{Data collapse of the ratio $f/h$ in the limit
$1/N_c\rightarrow 0$ for $\Delta=0.5E_M$.
Here the dimension of the Hamiltonian matrix takes values $N_c=12001, 8501,
8001, 7001, 6001, 5001, 4001, 3001, 2401, 2001$.}\label{fig:delta05}
\end{center}
\end{figure}

Let us calculate probability densities for finite values of
$\Delta/E_M$. The results for $\Delta/E_M=0.1$ are shown in
Fig.\ref{fig:prob2}(a), and we see that the results are similar to
those of $\Delta/E_M=0$. However, for a larger value $\Delta/E_M=1$
there is a second peak away from $r=0$, see Fig.\ref{fig:prob2}(b).

Fig.\ref{fig:invp2} displays the dimensionless inverse probability
density $f\Big(g,\frac{1}{N_c},\frac{\Delta}{E_M}\Big)$ for various
values of the coupling constant. The ratio between
$f\Big(g,\frac{1}{N_c},\frac{\Delta}{E_M}\Big)$ and
$h\Big(g,\frac{1}{N_c},\frac{\Delta}{E_M}\Big)$ approaches $1$ in
the limit $\frac{1}{N_c}\ll 1$, see the data collapse in
Fig.\ref{fig:delta05}.  This result suggests that, for $g\neq 0$,
the value of the exponent $\nu$ is given by Eq.(\ref{wavescal}),
{\it independent} of $\Delta$.

\section{Results of eigenenergies}

\subsection{$g>1/2$}

For $g>1/2$ no analytical result for eigenenergies exist. Our
numerical energy values of the state $|\Psi_0^{-1/2}\rangle$ at
$g=0.7$ and $\Delta=0$ are $E/E_M= -1.039 ,-1.063, -1.087, -1.114$
for $N_c= 4001, 6001,8501, 12001$. They diverge slowly in the limit
$N_c\rightarrow \infty$ . Similar results hold for $\Delta\neq 0$.
Since $\chi_{B}(r)\sim \frac{1}{r^{1/2}}$ for small $r$
(Eq.(\ref{wavescal})) the expectation value of the Coulomb potential
is
\begin{eqnarray}
E_C\sim \int_R^{\infty}dr r\Big( r^{-\frac{1}{2}}\frac{1}{r}
r^{-\frac{1}{2}}\Big)\sim \log(R).
\end{eqnarray}
It diverges slowly as $\log(R)$, consistent with our numerical
result.  Without the regularization parameter $R$ this energy
diverges.

\subsection{$g<1/2$}

Our numerical energy values of the state $|\Psi_0^{-1/2}\rangle$ for
$g=0.4$ converge fast as a function of $N_c$, in contrast to
$g>1/2$. Since its radial wavefunction $\chi_{B}(r)\sim
\frac{1}{r^{\nu}}$ with $\nu<1/2$ (Eq.(\ref{wavescal})) the
expectation value of the Coulomb potential is free of divergence
even at $R=0$.  No regularization parameter is thus needed when
$g<1/2$.

\begin{figure}[!hbpt]
\begin{center}
\includegraphics[width=0.8\textwidth]{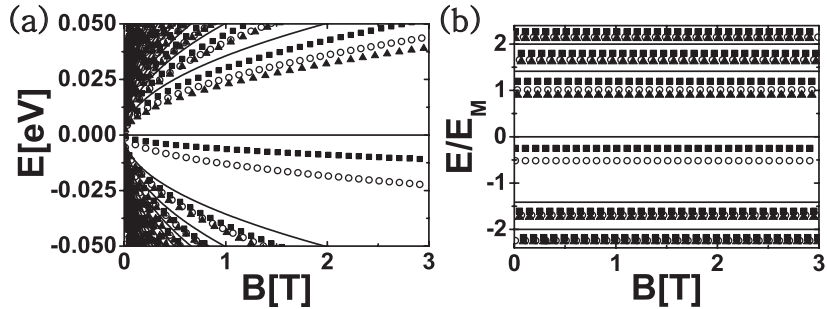}
\caption{Eigenenergies at $g=0.4$ for angular momenta  $J=-1/2$ ,
$J=-3/2$, and $J=1/2$, represented by circles, squares, and
triangles ($\Delta=0$ and  $N_c=8501$). Landau level energies in the
absence of the Coulomb impurity ($g=0$)  are represented by solid
lines.  In (a) eigenenergies measured in units of eV.  When they are
measured in units of $E_M$ the magnetic field dependence disappears
(b). }\label{fig:LL}
\end{center}
\end{figure}

\begin{figure}[!hbpt]
\begin{center}
\includegraphics[width=0.45\textwidth]{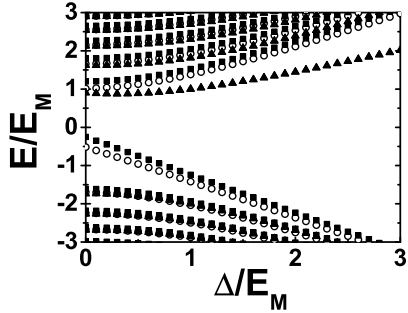}
\caption{  Eigenenergies as a function of  mass gap $\Delta$ at
$g=0.4$ and $N_c=8501$: $J=-1/2$ (circle), $J=-3/2$ (square), and
$J=1/2$ (triangle). }\label{fig:LL1}
\end{center}
\end{figure}

Fig.\ref{fig:LL}(a) displays eigenenergies as a function of $B$ when
$\Delta=0$. The same energies in units of $E_M$ are displayed in
Fig.\ref{fig:LL}(b). The independence of $E/E_M$ on $B$ reflects the
fact that the coupling constant $g$ is independent of $B$. Landau
energy levels of graphene are also shown for comparison. The
dimensionless eigenenergies for $\Delta\neq 0$ are displayed as a
function of $\Delta/E_M$ in Fig.\ref{fig:LL1}.  Using semiclassical
analysis for small $B$ Ho and Khalilov\cite{Ho} found that the first
positive energy $E_J$ less than $\Delta$ is given by
\begin{eqnarray}
\frac{E_J}{E_M}\approx\frac{\frac{\Delta}{E_M}+\frac{1}{2}\left(\frac{E_M}{\Delta}J
\right)}{\left[1+\frac{g^2}{\left[1+\sqrt{J^2-g^2}\right]^2}\right]^{1/2}}
\end{eqnarray}
(See Eq.(21) in Ref.\cite{Ho}).  From this expression we find that
$E_J/E_M\approx 9.534 $ for $g=0.4$, $J=-1/2$ and $\Delta/E_M=10$.
This value of energy agrees approximately  with the numerical value
$E_J/E_M\approx 9.571 $.  They also derived some exact solutions at
denumerable number of magnetic field values (not necessarily small),
and we will test our numerical results against them. According to
exact results of Eq.(39) in Ref.\cite{Ho} some negative energies $E_J$
less than $\Delta$ satisfy
\begin{eqnarray}
&&E_J=-\frac{\Delta}{2(\gamma+J+1/2)},\nonumber\\
&&E_J^2-\Delta^2=E^2_M(\gamma+1/2).
\end{eqnarray}
Solutions are
\begin{eqnarray}
\frac{E_J}{E_M}&=&-\frac{\sqrt{\gamma+1/2}}{\sqrt{-4(\gamma+J)(\gamma+J+1)}},\nonumber\\
\frac{\Delta}{E_M}&=&\frac{2(\gamma+J+1/2)\sqrt{\gamma+1/2}}{\sqrt{-4(\gamma+J)(\gamma+J+1)}},
\end{eqnarray}
where $\gamma=\sqrt{J^2-g^2}$, $J<0$, and $g<0.5$.  For $J=-1/2$ and
$g=0.4$ the solution is
$(\frac{\Delta}{E_M},\frac{E_J}{E_M})=(0.67,-1.118)$, in agreement
with the numerical result (0.67,-1.117) for $N_c=12001$.
%For $J=-3/2$ the analytical
%result is $(2.74,-3.08)$ and the numerical result for $N_c=8501$ is
%$(2.74,-2.97)$. We expect that the agreement between them would
%improve  for larger values of $N_c$.

\section {Conclusions and discussions}

\begin{figure}[!hbpt]
\begin{center}
\includegraphics[width=0.8\textwidth]{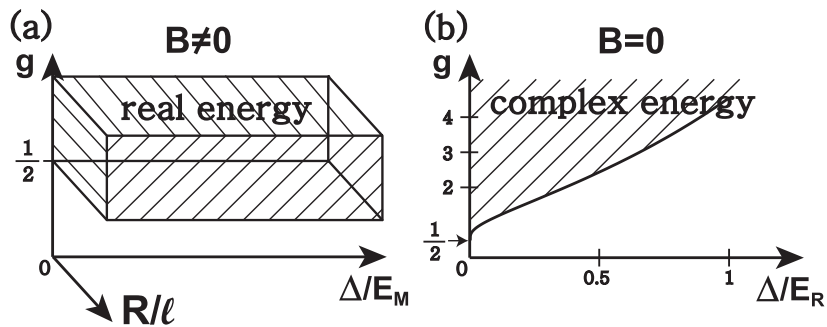}
\caption{ In the scratched regions the Coulomb potential must be
regularized. (a) $B\neq 0$  and (b) $B=0$.  Note that $E_M$ is the
energy scale of
 graphene LLs while $E_{R}$ is the energy scale of the regularization length.
 We have computed electronic wavefunctions in the scratched region at $B\neq0$.}\label{fig:prob10}
\end{center}
\end{figure}

We have explored eigenstates and eigenenergies of the Coulomb
problem in a magnetic field at finite values of renormalization
length $R$ and for different values of the coupling constant
$g>g_c$. As shown in Fig.\ref{fig:prob10} solutions are
qualitatively different from those of zero magnetic field since the
presence of a magnetic field prohibits complex energy solutions. Our
numerical results show that the inverse probability density of the
state $|\Psi_0^{-1/2}\rangle$ at $r=R$ is described by a scaling
function $f(g,\frac{1}{N_c},\frac{\Delta}{E_M})$, which exhibits a
significant dependence on $N_c$ or $R$. In the limit $N_c\rightarrow
\infty$ the wavefunction of its s-wave component behaves as
$\frac{1}{r^{\nu}}$ near $r=0$. We find that the exponent is
$\nu<1/2$ when $g<1/2$ and  $\nu=1/2$ when $g>1/2$, independent of
mass gap\cite{Gam0,Zh}.   We thus recover the previously known
results of the limit $R\rightarrow 0$, suggesting consistency of our
numerical method. The wavefunctions are {\it normalizable} for all
values of $g$.

In Ref.\cite {Gam1} an instability of many-body groundstate in
magnetic fields   is examined under the condition that an excited
energy coincides with  the Fermi energy, which is assumed to be
at $E_F=-\Delta$ (note also the value of the Fermi energy in
graphene can be tuned so it is not necessary always at
$E_F=-\Delta$). In our single electron problem in magnetic fields
this is not where the fall to the Coulomb center occurs.  Instead
the condition is  $E\rightarrow -\infty$\cite{Rei,Landau}.

In this paper we considered donor impurities. For acceptors or
antidots\cite{Weiss} we can use the transformation $V(r)\rightarrow
-V(r)$ with the eigenenergies $E\rightarrow -E$ (eigenstates are
unchanged when $\Delta =0$). It maybe worthwhile to investigate solutions in a
lattice model\cite{Zhu} instead of continuum models. Also the
coupling between $K$ and $K'$ valleys could provide an improved model.
An experimental test of our results may be performed by measuring
the transitions energies between the eigenenergies $E_N^{J}$.

\section*{Acknowledgments} This research was supported by Basic
Science Research Program through the National Research Foundation of
Korea(NRF) funded by the Ministry of Science, ICT $\&$ Future
Planning(MSIP) (NRF-2012R1A1A2001554). In addition this research was
supported by a Korea University Grant.


\begin{thebibliography}{00}
\bibitem{Rei} J. Reinhardt, W. Greiner, Rep. Prog. Phys. {\bf 40}, 219
(1977). Three dimensional Coulomb impurity  problem in the absence
of a magnetic field is reviewed.   Analytical solutions are
given.
\bibitem{Mac0} A. K. Geim
and A. H. MacDonald, Phys. Today {\bf 60}, 35 (2007).
\bibitem{Neto}
A. H. Castro Neto, F. Guinea, N. M. R. Peres, K. S. Novoselov, and A. K. Geim, Rev. Mod. Phys. {\bf 81}, 109 (2009).
\bibitem{Per}
V. M. Pereira, J. Nilsson, and A. H. Castro Neto,  Phys. Rev. Lett.
{\bf 99}, 166802 (2007); A. V. Shytov, M. I. Katsnelson, and L. S.
Levitov, Phys. Rev. Lett. {\bf 99}, 236801 (2007);  V. R. Khalilov
and C. L. Ho, Mod. Phys. Lett. A {\bf 13}, 615 (1998).
\bibitem{Gam0} O. V. Gamayun, E. V. Gorbar, and V. P. Gusynin,
Phys. Rev. B  {\bf 80}, 165429 (2009).
\bibitem{Mac} A. H. MacDonald and D. S. Ritchie, Phys. Rev. B  {\bf 33}, 8336
(1986).  In this paper Pade approximant is used to interpolate
between low and high magnetic field limits; M.Taut, J. Phys. A:
Math. Gen. {\bf 28}, 2081 (1995).  Analytical solutions are found at
a denumerably  set of magnetic fields.
\bibitem{Landau} L. D. Landau and E. M. Lifshitz, Quantum mechanics (3rd ed., Pergamon Press, Oxford, 1977).
\bibitem{com} The regularization parameter $R$ is the radius of the
charge: $V(r)=-\frac{e^2}{\epsilon \sqrt{r^2+R^2}}$  or
$V(r)=-\frac{e^2}{\epsilon R}$ for $r<R$ and $-\frac{e^2}{\epsilon
r}$ for $r>R$.
\bibitem{gia} G. Giavaras, P. A. Maksym, and M. Roy, J. Phys.: Condens. Matter {\bf
21}, 102201 (2009).
\bibitem{Rec}   A. Matulis and F.
M. Peeters, Phys. Rev. B  {\bf 77}, 115423 (2008);  P. S. Park, S.
C. Kim, and S. -R. Eric Yang, Phys. Rev. Lett.  {\bf 108}, 169701
(2012); S. C. Kim, J. W. Lee, and S. -R. Eric Yang, J. Phys.: Condens. Matter {\bf 24}, 495302 (2012).
\bibitem{Berry} M. V. Berry, Phys. Today {\bf 55}, 10 (2002).
\bibitem{Ben} C. Bender and S. Orszag, Advanced Mathematical Methods for
Scientists and Engineers  (McGraw Hill, New York, 1978).
\bibitem{Ho} C. L. Ho and V. R. Khalilov, Phys. Rev. A  {\bf 61}, 032104 (2000).
\bibitem{Zh} Y. Zhang, Y. Barlas, and K. Yang, Phys. Rev. B  {\bf 85}, 165423
(2012).
\bibitem{com01}  S. -R. Eric Yang, S. Mitra, A. H. MacDonald, and M. P. A. Fisher, J. Korean Phys. Soc.
{\bf 29}, S10 (1996).  The regularization length parameter $R$ is
related to wavevector $k$ through $R\sim 2\pi/k$.  A wavevector $k$
can be related to the average radius $\langle r\rangle$ through $k\ell^2=\langle r\rangle$.
Since state with a large  LL index $N_c$ has radius
$\langle r\rangle=\ell\sqrt{2N_c}$ we get $R\sim \ell\pi\sqrt{\frac{2}{N_c}}$.
\bibitem{Golden}  N. Goldenfeld,  Lectures on phase transitions and the renormalization
group  (Addison-Wesley, 1992).
\bibitem{Gam1} O. V. Gamayun, E. V. Gorbar, and V. P. Gusynin,
Phys. Rev. B  {\bf 83}, 235104 (2011);   a different type of
instability is investigated in P. S. Park, S. C. Kim, and S. -R. Eric
Yang, Phys. Rev.B {\bf 84} 085405 (2011).
\bibitem{Yosi}  D. Yoshioka, The
Quantum Hall Effect (Springer, Berlin, 1998).
\bibitem{Berry1} P. Recher, J. Nilsson, G. Burkard, and B.
Trauzettel, Phys. Rev. B {\bf 79}, 085407 (2009); M. V. Berry and R.
J. Mondragon, Proc. R. Soc. Lond. A  {\bf 412}, 53 (1987).
\bibitem{Weiss} P. S. Park, S. C. Kim, and S. -R. Eric Yang, J. Phys.: Condens. Matter  {\bf 22} 375302 (2010).
\bibitem{Zhu} W. Zhu, Z. Wang, Q. Shi, K. Y.
Szeto, J. Chen, and J. G. Hou, Phys. Rev. B {\bf 79}, 155430 (2009).


\end{thebibliography}
\end{document}